\documentclass{svjour3}                     
\usepackage{graphicx}
\usepackage{amsmath}
\usepackage{amssymb, amsfonts}
\begin{document}

\title{System plus reservoir approach to quantum Brownian motion of a rod-like particle}

\author{Z. Nasr \and F. Kheirandish }

\institute{Z. Nasr \at
              Department of Physics, Faculty of Science, University of Isfahan, Isfahan, Iran \\
              Tel.: +98-031-37934751\\
              Fax: +98-031-37932400\\
              \email{znmir2005@yahoo.com}           
           \and
           F. Kheirandish \at
              Department of Physics, Faculty of Science, University of Isfahan, Isfahan, Iran \\
              \email{fkheirandish@yahoo.com}}

\maketitle

\begin{abstract}
Quantum Brownian motion of a rod-like particle is investigated in the frame work of system plus reservoir model. The quantum mechanical and classical limit for both translational and rotational motions are discussed. Correlation functions, fluctuation-dissipation relations and mean squared values of translational and rotational motions are obtained.
\end{abstract}

\section{Introduction}
\label{intro}
\noindent
The theory of Brownian motion was described and formulated in its most elegant way by Albert Einstein in 1905 \cite{Einstein}. He derived a relationship between the diffusion coefficient of the particle and the viscosity of the fluid in which the particle is suspended. The general form of this relationship is known as the so-called fluctuation-dissipation theorem which is one of the building blocks of non-equilibrium statistical mechanics \cite{Kubo}. Not so long after the promising work of Einstein, others developed the classical theory of Brownian motion \cite{Uhlenbeck,Langevin,Smoluchowski}. Some authors analyzed Brownian motion from the purely mathematical point of view \cite{Wiener,Kolmogorov,Doob,Levy}. The concept of Brownian motion is not limited to the erratic motion of a particle in a fluid, it can be generalized to the fluctuating behavior of a system of interest interacting with its surroundings for instance molecular motion \cite{Nykypanchuk,Perrin}, a charged particle in an electromagnetic field \cite{Harko,Sandoval,Cobanera}, biological membranes \cite{Saffman,Frey,Jou}, motion in a spin bath \cite{Sinha}, polar molecules in external fields \cite{Waldron} and even modelling for stock market \cite{Meng} and so on.

There exists a variety of alternative but equivalent approaches to investigate Brownian motion. The most celebrated ones include probability distribution functions \cite{Hubbard1,Gardiner,Hubbard2} and Langevin method \cite{Patriarca,Weiss,Lahiri} for classical Brownian motion, and Feynman path integral \cite{Caldeira,Grabert,Zhang1,Zhang2,Fleming}, mainly concerned with deriving the master equation for both linear and non-linear couplings to the environment and white and colored noise and finally, operator Langevin equation \cite{Ford,Oppenheim} and stochastic methods \cite{Verdaguer} for quantum Brownian motion.
In the vast majority of the works mentioned above, the authors took advantage of system plus reservoir method of treating open systems.

In the present paper we describe the quantum Brownian motion of a rod-like particle which is interacting with a quantum reservoir and undergoes both translational and rotational motions. This kind of dynamics appears for example in the appealing subject of nano rod crystals and rigid rod-like macromolecules that have attracted remarkable attention especially in biological systems \cite{Okoshi,Riseman,Semenov}. In particular, rotational Brownian motion is involved in nuclear magnetic resonance (NMR) and electron paramagnetic resonance (EPR) interpretation of many physical phenomena \cite{Huertas,Steinhoff}.
However we follow the system plus reservoir approach to tackle the problem. There are two commonly used models for modeling a quantum reservoir or heat bath based on the properties of the reservoir. One can choose a reservoir consisting of non-interacting
quantum oscillators \cite{Hakim,Legget,Dekker1} if memory or non-Ohmic properties of the system are important or model the heat bath using appropriate scalar or vector fields \cite{Yurke,Dekker2} if Ohmic properties are considered. Here we follow the latter choice and couple the translational and rotational degrees of freedom of the particle with two independent vector fields.

The layout of the paper is as follows: in Sec. II the model is introduced and a total Lagrangian for the system and reservoir is given. In Sec. III
the equations of motion are derived, with two sub sections A and B devoted to translational and rotational motions respectively.
In Sec. IV, the correlation functions and mean values of linear and angular momentum are deduced. The mean square displacement and orientation in the Brownian time scales are calculated. Finally, we conclude in Sec. V.  
\section{The Model Lagrangian}
The motion of a rigid body is described in terms of the translational motion of its center of mass and the rotational motion of its orientation. One needs to have a convenient frame of reference to describe the motion, body-fixed frame or laboratory-fixed frame. We work in the laboratory-fixed frame.
Consider the special case of a long and thin rod (large aspect ratio) for which the rotational motion around the cylinder axis of symmetry is disregarded. This thin rod can be thought of as a prolate top since the inverse of its inertial momentum around its axis is much larger than the perpendicular components. The angular velocity $\mathbf{\Omega}$ is thought of as being perpendicular to the cylinder axis of symmetry.

The surrounding medium exerts a fluctuating force as well as a fluctuating torque on the rod. For Ohmic damping the medium can be modeled by two independent vector fields, coupled to translational and orientational degrees of freedom of the rod. The Lagrangian of the total system, rod plus medium reads
\begin{eqnarray}\label{Lagrangian}
 L &=& \frac{1}{24}Ml^{2}\mathbf{\Omega}^{2}+\frac{1}{2}M\mathbf{V}^{2}+
 \frac{1}{2}\int_{0}^{\infty}dx\,\left[(c^{-1}\partial_{t}\boldsymbol\varphi(x,t))^{2}-(\partial_{x}\boldsymbol\varphi(x,t))^{2}\right]\nonumber \\
&&+\frac{1}{2}\int_{0}^{\infty}dx\,\left[(c^{-1}\partial_{t}\boldsymbol\psi(x,t))^{2}-(\partial_{x}\mathbf{\boldsymbol\psi}(x,t))^{2}\right]\nonumber\\
&&-\alpha\,\mathbf{\Omega}\cdot\boldsymbol\varphi(0,t)-\mathbf{V}\cdot\mathbf{\Gamma}'_{f}\cdot\boldsymbol\psi(x,t),
\end{eqnarray}
in the above equations $\boldsymbol\varphi(x,t)$ and $\boldsymbol\psi(x,t)$ are the vector fields describing the medium, $\alpha$ is the coupling constant between $\boldsymbol\varphi(x,t)$ and $\mathbf{\Omega}$. The friction tensor $\mathbf{\Gamma}'_{f}$ couples $\boldsymbol\psi(x,t)$ with the translational velocity.

It is most convenient to decompose the fields in body frame where parallel and perpendicular means parallel to the symmetry axis of the rod and perpendicular to this axis, see Fig. 1. So we can write
\begin{equation}\label{coupling}
\mathbf{\Gamma}'_{f}=\gamma'_{\parallel}\hat{u}\hat{u}+\gamma'_{\perp}(1-\hat{u}\hat{u}),
\qquad \mathbf{V}\cdot\mathbf{\Gamma}'_{f}=\gamma'_{\parallel}\mathbf{V}_{\parallel}+\gamma'_{\perp}\mathbf{V}_{\perp},
\end{equation}
where $\hat{u}\hat{u}$ is the dyadic tensor in the direction of the rod. Now if we rewrite the Lagrangian in terms of these generalized variables we get
\begin{eqnarray}
L &=&\frac{1}{24}Ml^{2}\mathbf{\Omega}^{2}+\frac{1}{2}M(\mathbf{V}_{\parallel}^{2}+\mathbf{V}_{\perp}^{2})+
\frac{1}{2}\int_{-\infty}^{+\infty}dx\,\left\{c^{-2}(\partial_{t}\boldsymbol\varphi)^{2}-(\partial_{x}\boldsymbol\varphi)^{2}
\right\}\nonumber\\
&& +\frac{1}{2}\int_{-\infty}^{+\infty}dx\,\left\{c^{-2}(\partial_{t}\boldsymbol\psi_{\parallel})^{2}-(\partial_{x}\boldsymbol\psi_{\parallel})^{2}
+c^{-2}(\partial_{t}\boldsymbol\psi_{\perp})^{2}-(\partial_{x}\boldsymbol\psi_{\perp})^{2}\right\}\nonumber\\
&&-\alpha\,\mathbf{\Omega}\cdot\boldsymbol\varphi(0,t)-\gamma'_{\parallel}\mathbf{V}_{\parallel}\cdot\boldsymbol\psi_{\parallel}(0,t)-
\gamma'_{\perp}\mathbf{V}_{\perp}\cdot\boldsymbol\psi_{\perp}(0,t).
\end{eqnarray}
The generalized momenta are defined by
\begin{eqnarray}\label{momenta}
  \boldsymbol\pi_\varphi &=& c^{-2}\dot{\boldsymbol\varphi}, \\
  \boldsymbol\pi_{\boldsymbol\psi_{\parallel}} &=& c^{-2}\dot{\boldsymbol\psi}_{\parallel},  \\
   \boldsymbol\pi_{\boldsymbol\psi_{\perp}} &=& c^{-2}\dot{\boldsymbol\psi}_{\perp}, \\
  \mathbf{P}_{\parallel} &=& M \mathbf{V}_{\parallel}-\gamma'_{\parallel}\boldsymbol\psi_{\parallel} (0,t),  \\
  \mathbf{P}_{\perp} &=& M \mathbf{V}_{\perp}-\gamma'_{\parallel}\boldsymbol\psi_{\perp} (0,t), \\
  \mathbf{P}_{\Omega} &=& I \boldsymbol\Omega -\alpha \boldsymbol\varphi (0,t),\,\,\,\,\,I=\frac{Ml^2}{12}.
\end{eqnarray}
The corresponding Hamiltonian has the following minimal coupling structure
\begin{eqnarray}\label{Hamiltonian}
  H &=& \frac{[\mathbf{P}_{\parallel}+\gamma'_{\parallel}\,\boldsymbol\psi_{\parallel}(0,t)]^2}{2M}+
  \frac{[\mathbf{P}_{\perp}+\gamma'_{\perp}\,\boldsymbol\psi_{\perp}(0,t)]^2}{2M}+\frac{[\mathbf{P}_{\Omega}+\alpha\,\boldsymbol\varphi (0,t)]^2}{2I} \\
   &+& \frac{1}{2}\int dx\,\{c^2 \boldsymbol\pi_\varphi^2 +(\partial_x \boldsymbol\varphi)^2\}+
    \frac{1}{2}\int dx\,\{c^2 \boldsymbol\pi_{\varphi_\parallel}^2 +(\partial_x \boldsymbol\psi_{\parallel})^2\} \\
   &+& \frac{1}{2}\int dx\,\{c^2 \boldsymbol\pi_{\varphi_\perp}^2 +(\partial_x \boldsymbol\psi_{\perp})^2\}.
\end{eqnarray}
To quantize the system we impose the following equal-time commutation relations
\begin{eqnarray}\label{quantization}
  && [\boldsymbol\varphi (x,t), \boldsymbol\pi_{\varphi} (x',t)] = i\hbar\,\delta (x-x')\mathbb{I} \\
  && [\boldsymbol\psi_{\perp} (x,t), \boldsymbol\pi_{\perp} (x',t)] = i\hbar\,\delta (x-x')\mathbb{I} \\
  && [\boldsymbol\psi_{\parallel} (x,t), \boldsymbol\pi_{\parallel} (x',t)] = i\hbar\,\delta(x-x')\mathbb{I} \\
  && [\mathbf{R}_{\parallel},\mathbf{P}_{\parallel}] = i\hbar\mathbb{I} \\
  && [\mathbf{R}_{\perp},\mathbf{P}_{\perp}] = i\hbar\mathbb{I} \\
  && [\mathbf{\Omega},\mathbf{P}_{\Omega}] = i\hbar\mathbb{I},
\end{eqnarray}
where $\mathbb{I}=\mbox{diag}(1,1,1)$ is the unit matrix.
\section{Equations Of Motion}
\subsection{Translational Motion}
From Hamiltonian (\ref{Hamiltonian}) and Heisenberg equations for the translational momentum we find
\begin{equation}\label{M1}
\frac{MdV_{\parallel(\perp)}}{dt}=\gamma'_{\parallel(\perp)}\mathbf{\dot{\psi}}_{\parallel(\perp)}(0,t).
\end{equation}
The same equations for the vector fields $\boldsymbol\psi_{\parallel(\perp)}(x,t)$ yield
\begin{equation}\label{M2}
(\frac{1}{c^2}\partial_{t}^{2}-\partial_{x}^{2})\boldsymbol\psi_{\parallel(\perp)}(x,t)=-\gamma'_{\parallel(\perp)}\mathbf{V}_{\parallel(\perp)}\delta(x),
\end{equation}
with the solution
\begin{equation}
\boldsymbol\psi_{\parallel(\perp)}(x,t)=\boldsymbol\psi^{h}_{\parallel(\perp)}(x,t)-\frac{\gamma'_{\parallel(\perp)}c}{2}
\int_{-\infty}^{t-\frac{|x|}{c}}dt'\,\mathbf{V}_{\parallel(\perp)}(t').
\end{equation}
The field $\boldsymbol\psi^{h}(x,t)$ is the solution to the homogenous equation that can be expressed in terms of
creation and annihilation operators as
\begin{eqnarray}
\boldsymbol\psi_{\parallel}^{h}(x,t) &=& \int\limits_{-\infty}^{+\infty}dk\,\sqrt{\frac{\hbar c^{2}}{4\pi\omega_{k}}}\,\hat{u}
\left\{\hat{b}_{\parallel}(k)e^{ikx-i\omega_{k}t}+\hat{b}^{\dag}_{\parallel}(k)e^{-ikx+i\omega_{k}t}\right\}, \nonumber\\
\boldsymbol\psi_{\perp}^{h}(x,t) &=& \int\limits_{-\infty}^{+\infty}dk\,\sqrt{\frac{\hbar c^{2}}{4\pi\omega_{k}}}\sum_{\lambda=1}^{2}\hat{\mathbf{e}}_{\lambda}
\left\{\hat{b}_{\perp}(k,\lambda)e^{ikx-i\omega_{k}t}+\hat{b}^{\dag}_{\perp}(k,\lambda)e^{-ikx+i\omega_{k}t}\right\}.\nonumber\\
\end{eqnarray}
the operators $\hat{b}_{\parallel}$ and $\hat{b}_{\perp}$ satisfy the bosonic commutation relations
\begin{equation}
[\hat{b}_{\parallel}(k),\hat{b}^{\dag}_{\parallel}(k')]=\delta(k-k'),\qquad
[\hat{b}_{\perp}(k,\lambda),\hat{b}^{\dag}_{\perp}(k',\lambda')]=\delta_{\lambda\lambda'}\delta(k-k').
\end{equation}
with all other commutators being zero. The transverse polarization unit vectors satisfy $\hat{\mathbf{e}}_{\lambda}\cdot\hat{\mathbf{e}}_{\lambda'}=\delta_{\lambda\lambda'}$, and with $\hat{u}$ make an orthogonal frame.
Now that the explicit form of the field is known, the translational velocity can be fully determined
\begin{equation}\label{M4}
\frac{d\mathbf{p}}{dt}+\frac{1}{M}\mathbf{\Gamma}_{f}\cdot\mathbf{p}=\mathbf{f}(t),
\end{equation}
this is the well-known Langevin equation in which $\mathbf{f}(t)$ and $\mathbf{\Gamma}_{f}$ are defined as
\begin{eqnarray}
\mathbf{\Gamma}_{f}&=&\gamma_{\parallel}\hat{u}\hat{u}+\gamma_{\perp}(1-\hat{u}\hat{u}),\quad \gamma_{\parallel(\perp)}=\frac{\gamma'^{2}_{\parallel(\perp)}c}{2}\nonumber\\
\mathbf{\mathbf{f}}(t)&=&\left[\sqrt{\frac{2\gamma_{\parallel}}{c}}\hat{u}\hat{u}+\sqrt{\frac{2\gamma_{\perp}}{c}}
(1-\hat{u}\hat{u})\right]\cdot\dot{\boldsymbol\psi}_{h}(0,t).
\end{eqnarray}
Since the noise field is a linear expression in creation and annihilation operators its expectation value is zero as expected. Eq. (\ref{M1}) can be integrated to give the translational momentum. For this purpose note that the friction tensor $\mathbf{\Gamma}_{f}$ has the property
\begin{equation}
\mathbf{\Gamma}_{f}^{n}=\gamma_{\parallel}^{n}\hat{u}\hat{u}+\gamma_{\perp}^{n}(1-\hat{u}\hat{u})
\Longrightarrow \mathbf{\Gamma}_{f}^{\frac{1}{2}}=\sqrt{\gamma_{\parallel}}\hat{u}\hat{u}+\sqrt{\gamma_{\perp}}(1-\hat{u}\hat{u}),
\end{equation}
using this property the Green function
$$\mathbf{G}(t-t')=e^{-\frac{\mathbf{\Gamma}_{f}(t-t')}{M}},$$
takes the form
$$ e^{-\frac{\mathbf{\Gamma}_{f}(t-t')}{M}}=e^{-\frac{\gamma_{\parallel}(t-t')}{M}}\hat{u}\hat{u}
+e^{-\frac{\gamma_{\perp}(t-t')}{M}}(1-\hat{u}\hat{u}),$$
leading to the following equations for the transverse and parallel components of the momentum of the rod
\begin{eqnarray}
\mathbf{p}_{\parallel}(t)&=&e^{-\frac{\gamma_{\parallel}t}{M}}\mathbf{p}_{\parallel}(0)+
\int_{0}^{t}dt'\,e^{-\frac{\gamma_{\parallel}(t-t')}{M}}\mathbf{f}_{\parallel}(t'),\nonumber \\
\mathbf{p}_{\perp}(t) &=& e^{-\frac{\gamma_{\perp}t}{M}}\mathbf{p}_{\perp}(0)+
\int_{0}^{t}dt'\,e^{-\frac{\gamma_{\perp}(t-t')}{M}}\mathbf{f}_{\perp}(t'),
\end{eqnarray}
where
\begin{eqnarray}
\mathbf{f}_{\parallel}(t)&=&i\sqrt{2\gamma_{\parallel}c}\int\limits_{-\infty}^{+\infty}dk\,
\sqrt{\frac{\hbar\omega_{k}}{4\pi}}\,\hat{u}\left\{
\hat{b}^{\dag}_{\parallel}(k)e^{i\omega_{k}t}-\hat{b}_{\parallel}(k)e^{-i\omega_{k}t}\right\},\nonumber\\
\mathbf{f}_{\perp}(t)&=& i\sqrt{2\gamma_{\perp}c}\int\limits_{-\infty}^{+\infty}dk\,
\sqrt{\frac{\hbar\omega_{k}}{4\pi}}\sum_{\lambda=1}^{2}\hat{\mathbf{e}}_{\lambda}\left\{
\hat{b}^{\dag}_{\perp}(k)e^{i\omega_{k}t}-\hat{b}_{\perp}(k)e^{-i\omega_{k}t}\right\}.
\end{eqnarray}
\subsection{Rotational Motion}
The angular velocity of the rod is perpendicular to the rod orientation. Using
the Lagrangian (\ref{Lagrangian}), the equation of motion for the angular velocity yields
\begin{equation}\label{M7}
\frac{1}{12}Ml^{2}\frac{d\mathbf{\Omega}}{dt}=\alpha\mathbf{\dot{\varphi}}_{h}(0,t),
\end{equation}
also for the vector field $\boldsymbol\varphi_{h}(x,t)$ we have
\begin{equation}\label{M8}
\left(\frac{1}{c^{2}}\partial_{t}^{2}-\partial_{x}^{2}\right)\boldsymbol\varphi_{h}(x,t)=-\alpha\delta(x)\mathbf{\Omega}(t).
\end{equation}
Following the lines of the previous section
\begin{equation}\label{M9}
\boldsymbol\varphi(x,t)=\boldsymbol\varphi_{h}(x,t)-\frac{\alpha c}{2}\int_{-\infty}^{t-\frac{|x|}{c}}dt'\,\mathbf{\Omega}(t'),
\end{equation}
where
\begin{equation}\label{M10}
\boldsymbol\varphi_{h}(x,t)=\int\limits_{-\infty}^{+\infty}dk\,\sqrt{\frac{\hbar c^{2}}{4\pi\omega_{k}}}
\sum_{\lambda=1}^{2}\hat{\mathbf{e}}_{\lambda}\left[\mathbf{a}(k,\lambda)e^{ikx-i\omega_{k}t}+
\mathbf{a}^{\dag}(k,\lambda)e^{-ikx+i\omega_{k}t}\right].
\end{equation}
The Langevin equation for the angular velocity is
\begin{equation}\label{M11}
\frac{1}{12}Ml^{2}\frac{d\mathbf{\Omega}(t)}{dt}+\mathbf{\gamma}_{r}\mathbf{\Omega}(t)=\mathbf{T}(t),
\end{equation}
where $\mathbf{\gamma}_{r}=\alpha^{2}c/2$ and
\begin{eqnarray}
\mathbf{T}(t) &=& \alpha\,\dot{\boldsymbol\psi}_{h}(0,t)\nonumber\\
&=& i\alpha\int\limits_{-\infty}^{+\infty}dk\,\sqrt{\frac{\hbar c^{2}\omega_{k}}{4\pi}}\sum_{\lambda=1}^{2}\hat{\mathbf{e}}_{\lambda}
\left[\hat{a}^{\dag}(k,\lambda)e^{i\omega_{k}t}-\hat{a}(k,\lambda)e^{-i\omega_{k}t}\right].
\end{eqnarray}
Integrating Eq. (\ref{M11}) results in the angular velocity
\begin{equation}\label{M12}
\mathbf{\Omega}(t)=e^{-\frac{12\gamma_{r}t}{Ml^{2}}}\mathbf{\Omega}(0)+
\frac{12}{Ml^{2}}\int_{0}^{t'}e^{-\frac{12}{Ml^{2}}(t-t')}\mathbf{T}(t').
\end{equation}
The explicit form of the angular velocity  in the long time limit can be obtained in terms of the creation and annihilation operators as
\begin{equation}\label{M13}
\mathbf{\Omega}(t)=\frac{1}{I}\int_{-\infty}^{+\infty}dk\,\sqrt{\frac{\hbar c^2 \omega_{k}}{4\pi}}\sum_{\lambda=1}^{2}\mathbf{\hat{e}}_{\lambda}
\left[\frac{\hat{a}^{\dag}(k,\lambda)e^{i\omega_{k}t}}{\omega_{k}-i\gamma_{r}}+h.c.\right],
\end{equation}
likewise, the angular momentum $ \mathbf{J}=I\mathbf{\Omega}$ can be written as
\begin{equation}\label{M14}
\mathbf{J}(t)=\int_{-\infty}^{+\infty}dk\,\sqrt{\frac{\hbar c^2 \omega_{k}}{4\pi}}\sum_{\lambda=1}^{2}\mathbf{\hat{e}}_{\lambda}
\left[\frac{\hat{a}^{\dag}(k,\lambda)e^{i\omega_{k}t}}{\omega_{k}-i\gamma_{r}}+h.c.\right].
\end{equation}
\section{Correlation Functions}
Due to the linearity of the coupling between the system and the reservoir the noise force and torque
obey Gaussian statistics. However their auto correlation functions play a significant role
in determining how the environment affects the system. The parallel and perpendicular components of the
fluctuating force and the torque were obtained as
\begin{eqnarray}
f_{\parallel}(t) &=& i\sqrt{\frac{2\hbar c\gamma_{\parallel}}{4\pi}}\int_{-\infty}^{+\infty}dk\,\sqrt{\omega_{k}}
\left\{\hat{b}^{\dag}_{\parallel}(k)e^{i\omega_{k}t}-\hat{b}_{k}e^{-i\omega_{k}t}\right\},\nonumber\\
\mathbf{f}_{\perp}(t) &=& i\sqrt{\frac{2\hbar c\gamma_{\perp}}{4\pi}}\int_{-\infty}^{+\infty}dk\,
\sqrt{\omega_{k}}\sum_{\lambda=1}^{2}\hat{\mathbf{e}}_{\lambda}
\left\{\hat{b}^{\dag}_{\perp}(k,\lambda)e^{i\omega_{k}t}-\hat{b}_{k,\lambda}e^{-i\omega_{k}t}\right\},\nonumber\\
\mathbf{T}(t) &=& ic\alpha\sqrt{\frac{\hbar}{4\pi}}\int_{-\infty}^{+\infty}dk\,\sqrt{\omega_{k}}
\sum_{\lambda=1}^{2}\hat{\mathbf{e}}_{\lambda}
\left\{\mathbf{a}^{\dag}(k,\lambda)e^{i\omega_{k}t}-\mathbf{a}(k,\lambda)e^{-i\omega_{k}t}\right\}.
\end{eqnarray}
Regarding the commutation properties of $\hat{b}^{(\dag)}_{\parallel}$ and $\hat{b}^{(\dag)}_{\perp}$, for the correlation between $\mathbf{f}_{\parallel}$ and $\mathbf{f}_{\perp}$ we find
\begin{equation}\label{}
  \langle\mathbf{f}_{\parallel}(t)\cdot\mathbf{f}_{\perp}(t')\rangle=0,
\end{equation}
the symmetric correlations for each component is
\begin{eqnarray}
\frac{1}{2}\langle\mathbf{f}_{\parallel}(t)\cdot\mathbf{f}_{\parallel}(t')\rangle_{\mbox{sym}} &=& \frac{\hbar c\gamma_{\parallel}}{2\pi} \int\limits_{-\infty}^{+\infty}dk\,\omega_{k}\coth(\frac{\hbar\omega_{k}}{2K_{B}T})\cos \omega_{k}(t-t'),\nonumber\\
\frac{1}{2}\langle\mathbf{f}_{\perp}(t)\cdot\mathbf{f}_{\perp}(t')\rangle_{\mbox{sym}} &=& \frac{2\hbar c\gamma_{\perp}}{2\pi} \int\limits_{-\infty}^{+\infty}dk\,\omega_{k}\coth(\frac{\hbar\omega_{k}}{2K_{B}T})\cos \omega_{k}(t-t'),
\end{eqnarray}
where $\langle a(t)b(t')\rangle_{\mbox{sym}}=\langle a(t)b(t')+b(t')a(t)\rangle/2$. In high-temperature regime (classical limit) the above equations turn into
\begin{eqnarray}
\langle\mathbf{f}_{\parallel}(t)\cdot\mathbf{f}_{\parallel}(t')&=& 2K_{B}T\gamma_{\parallel}\delta(t-t')\nonumber\\
\langle\mathbf{f}_{\perp}(t)\cdot\mathbf{f}_{\perp}(t')&=& 4K_{B}T\gamma_{\perp}\delta(t-t'),
\end{eqnarray}
and we recover the fluctuation-dissipation relation. The terms $2K_{B}T\gamma_{\parallel}$ and $4K_{B}T\gamma_{\perp}$ are the parallel and perpendicular
fluctuation strengths respectively. It is seen that the perpendicular fluctuating strength is twice
its parallel counter part.

Using the expression given for the torque, the correlation at different times is obtained as
\begin{equation}\label{N1}
\frac{1}{2}\langle\mathbf{T}(t)\cdot\mathbf{T}(t')\rangle_{\mbox{sym}}
=\frac{c\hbar\gamma_{r}}{\pi}\int\limits_{-\infty}^{+\infty}dk\,\omega_{k}\coth(\frac{\hbar\omega_{k}}{2K_{B}T})
\cos \omega_{k}(t-t'),
\end{equation}
and in the classical limit we find
\begin{equation}\label{N2}
\langle\mathbf{T}(t)\cdot\mathbf{T}(t')\rangle=2K_{B}T\gamma_{r}\delta(t-t'),
\end{equation}
so the rotational fluctuating strength is $2K_{B}T\gamma_{r}$.

Now that the force and torque correlations are known one can go further and calculate the correlations
between the desired variables, i.e translational and angular velocity. Due to the fact that the long time
behaviour of these variables are of interest, the terms having the decaying exponential prefactor do not
contribute in correlations, after performing the required manipulations it is found that
\begin{eqnarray}
\lim_{t\longrightarrow\infty}\langle\mathbf{p}_{\parallel}(t)\cdot\mathbf{p}_{\parallel}(t)\rangle &=& MK_{B}T,
\nonumber \\
\lim_{t\longrightarrow\infty}\langle\mathbf{p}_{\perp}(t)\cdot\mathbf{p}_{\perp}(t)\rangle &=&
2MK_{B}T,
\end{eqnarray}
and
\begin{equation}\label{N3}
\lim_{t\longrightarrow\infty}\langle\mathbf{\Omega}(t)\cdot\mathbf{\Omega}(t)\rangle=\frac{12}{Ml^{2}}K_{B}T.
\end{equation}
One can use the obtained mean values for the parallel and perpendicular momenta and the angular velocity to calculate
the mean energy of the system
\begin{equation}\label{Ener}
\langle E \rangle=\frac{1}{2}I\langle\Omega^{2}\rangle+\frac{\langle P_{\parallel}^{2}\rangle}{2M}+\frac{\langle P_{\perp}^{2}\rangle}{2M}.
\end{equation}
Substituting the obtained values in (\ref{Ener}), the total energy of the system reads
\begin{equation}
\langle E \rangle=\frac{1}{2M}\{MK_{B}T+2MK_{B}T\}+\frac{1}{2I}2IK_{B}T=\frac{5}{2}K_{B}T,
\end{equation}
which equals the classical energy of a molecule with five degrees of freedom. However for quantum mechanical case, due to the divergence of the integrals of the type
\begin{equation}
\int_{-\infty}^{+\infty}dk\,\frac{\omega_{k}\coth{\frac{\hbar\omega_{k}}{2K_{B}T}}}{\omega_{k}^{2}+\gamma^{2}},
\end{equation}
the energy of the system is divergent as has been stated in the literature \cite{Philbin} about the unphysical results of ohmic approximation in quantum mechanics. Usually one can be less strict about these divergencies and circumvent the problem by introducing a cut-off frequency in the upper limit of the integral.
\subsection{Mean-Square Displacement}
In the Brownian time scale for which the momentum of the Brownian particle relaxes due to the friction
with its environment, inertial forces can be neglected so the fore $\frac{d\mathbf{p}}{dt}$
can be omitted from the following equation of motion \cite{Gompper}
\begin{equation}
\frac{d\mathbf{p}}{dt}=-\mathbf{\Gamma}_{f}\cdot\mathbf{p}+\mathbf{f}(t),
\end{equation}
hence one can write
\begin{equation}\label{}
  \frac{d\mathbf{r}}{dt}=\frac{1}{M}\mathbf{\Gamma}_{f}^{-1}\cdot\mathbf{f}(t).
\end{equation}
According to the properties of the friction tensor $\mathbf{\Gamma}_{f}$ we have
\begin{equation}\label{P1}
\frac{d\mathbf{r}}{dt}=\frac{1}{\gamma_{\parallel}}\mathbf{f}_{\parallel}(t)+\frac{1}{\gamma_{\perp}}\mathbf{f}_{\perp}(t),
\end{equation}
therefore,
\begin{equation}\label{P2}
\mathbf{r}(t)=\mathbf{r}(0)+\int_{0}^{t}dt'\,\left[\frac{1}{\gamma_{\parallel}}\mathbf{f}_{\parallel}(t')+\frac{1}{\gamma_{\perp}}\mathbf{f}_{\perp}(t')\right].
\end{equation}
The mean square displacement, i.e $\langle|\mathbf{r}(t)-\mathbf{r}(0)|^{2}\rangle $ is found as
\begin{equation}\label{P3}
\langle|\mathbf{r}(t)-\mathbf{r}(0)|^{2}\rangle=2K_{B}Tt\left(\frac{1}{\gamma_{\parallel}}+\frac{2}{\gamma_{\perp}}\right),
\end{equation}
which is consistent with the known results for Einstein translational diffusion coefficients
$D_{\parallel}=\frac{K_{B}T}{\gamma_{\parallel}}$ and $D_{\perp}=\frac{K_{B}T}{\gamma_{\perp}}.$
\subsection{Orientation of the rod}
The orientation of the rod is a unit vector which is stationary in the body frame but in the laboratory frame it changes with time according to the following relation
\begin{equation}
\frac{d\mathbf{{\hat{u}}}(t)}{dt}=\mathbf{\Omega}(t)\times\mathbf{\hat{u}}(t),
\end{equation}
in order to find $ \mathbf{\hat{u}}(t) $ in the lab frame, we make use of the Euler parametrization which is associated with three successive
rotations through the angles $\alpha,\beta$ and $\gamma$. The rotation matrix is given by
\begin{equation}\label{R}
R=
\left(
  \begin{array}{ccc}
\cos\alpha\cos\beta\cos\gamma-\sin\alpha\sin\gamma & -\cos\alpha\cos\beta\sin\gamma-\sin\alpha\cos\gamma & -\cos\alpha\sin\beta \\
\sin\alpha\cos\beta\cos\gamma+\cos\alpha\sin\gamma & -\sin\alpha\cos\beta\sin\gamma+\cos\alpha\cos\gamma & -\cos\alpha\sin\beta \\
\sin\beta\cos\gamma & -\sin\beta\sin\gamma & \cos\beta\\
  \end{array}
\right).
\end{equation}
The angles $\alpha,\beta,\gamma$ are functions of time. Now if we denote the unit vectors in the body frame by $ \hat{\mathbf{e}_{1}},\hat{\mathbf{e}_{2}}$ and $\hat{\mathbf{e}_{3}}$, we can assume that $\hat{\mathbf{e}_{3}}$ is $\mathbf{\hat{u}}(t)$
in the lab frame showing the orientation of the rod. The unit vectors in the lab frame are the conventional $ \mathbf{\hat{i}},\mathbf{\hat{j}},\mathbf{\hat{k}}.$ Using the rotation matrix $R$ the relation between the two frames is obtained
\begin{eqnarray}
\mathbf{\hat{e_{1}}} &=& (\cos\alpha\cos\beta\cos\gamma-\sin\alpha\sin\gamma)\mathbf{\hat{i}}- (\cos\alpha\cos\beta\sin\gamma+\sin\alpha\cos\gamma)\mathbf{\hat{j}}\nonumber\\
&-& \cos\alpha\sin\beta \mathbf{\hat{k}}, \nonumber\\
\mathbf{\hat{e_{2}}} &=& (\sin\alpha\cos\beta\cos\gamma+\cos\alpha\sin\gamma)\mathbf{\hat{i}}-(\sin\alpha\cos\beta\sin\gamma-\cos\alpha\cos\gamma)\mathbf{\hat{j}}\nonumber\\
&-& \sin\alpha\sin\beta\mathbf{\hat{k}}, \nonumber\\
\mathbf{\hat{e_{3}}} &=& \sin\beta\cos\gamma\mathbf{\hat{i}}-\sin\beta\sin\gamma\mathbf{\hat{j}}+\cos\beta\mathbf{\hat{k}}.
\end{eqnarray}
Therefore,
\begin{equation}\label{U3}
 \mathbf{\hat{e_{3}}}
\equiv\mathbf{\hat{u}}(t)=\sin\beta(t)\cos\gamma(t)\mathbf{\hat{i}}-\sin\beta(t)\sin\gamma(t)\mathbf{\hat{j}}+\cos\beta(t)\mathbf{\hat{k}}.
\end{equation}
In the previous sections the angular velocity in the body frame was obtained as
\begin{equation}\label{Q1}
\mathbf{\Omega}=\frac{1}{I}\int_{-\infty}^{+\infty}dk\,\sqrt{\frac{\hbar c^{2}\omega_{k}}{4\pi}}\sum_{\lambda=1}^{2}\mathbf{\hat{e}_{\lambda}}
\left[\frac{\mathbf{a}^{\dag}(k,\lambda)e^{i\omega_{k}t}}{\omega_{k}-i\frac{\gamma_{r}}{I}}+h.c.\right]
\equiv\mathbf{\Omega}_{1}(t)\mathbf{\hat{e}}_{1}+\mathbf{\Omega}_{2}(t)\mathbf{\hat{e}}_{2},
\end{equation}
which in the lab frame is
\begin{eqnarray}
\mathbf{\Omega}(t)&=&\left[(\mathbf{\Omega}_{1}\cos\alpha+\mathbf{\Omega}_{2}\sin\alpha)\cos\beta\cos\gamma+
(\mathbf{\Omega}_{2}\cos\alpha-\mathbf{\Omega}_{1}\sin\alpha)\sin\gamma\right]\mathbf{\hat{i}} \nonumber\\
&+&\left[(\mathbf{\Omega}_{2}\cos\alpha-\mathbf{\Omega}_{1}\sin\alpha(t))\cos\gamma-
 (\mathbf{\Omega}_{1}\cos\alpha+\mathbf{\Omega}_{2}\sin\alpha)\cos\beta\sin\gamma\right]\mathbf{\hat{j}} \nonumber\\
&-& (\mathbf{\Omega}_{1}\cos\alpha+\mathbf{\Omega}_{2})\sin\beta\mathbf{\hat{k}},
\end{eqnarray}
where for notational convenience explicit time dependence of parameters has been ignored. Now recall the Langevin equation for $\mathbf{\Omega}(t)$
\begin{equation}
  \frac{d\mathbf{J}}{dt}=-\gamma_{r}\mathbf{\Omega}(t)+\mathbf{T}(t),
\end{equation}
as it was justified before (for mean square displacement) in the Brownian time scale one can neglect the time derivative
of angular momentum so that the angular velocity can be written as
\begin{equation}\label{Omega}
  \mathbf{\Omega}(t)=\frac{1}{\gamma_{r}}\mathbf{T}(t).
\end{equation}
The time derivative of the unit vector $\mathbf{\hat{u}}(t)$ takes the following form
\begin{equation}\label{udot}
  \frac{d\mathbf{\hat{u}}(t)}{dt}=\frac{1}{\gamma_{r}}\mathbf{T}(t)\times\mathbf{\hat{u}}(t),
\end{equation}
introducing an antisymmetric matrix
\begin{equation}\label{F}
 \mathbf{F}(t)=\left(
     \begin{array}{ccc}
       0 & -T_{3}(t) & T_{2}(t) \\
       T_{3}(t) & 0 & -T_{1}(t) \\
       -T_{2}(t) & T_{1}(t) & 0 \\
     \end{array}
   \right),
\end{equation}
equation (\ref{udot}) can be rewritten as
\begin{equation}\label{udotf}
  \frac{d\mathbf{\hat{u}}(t)}{dt}=\mathbf{F}(t)\cdot\mathbf{\hat{u}}(t).
\end{equation}
The matrix $\mathbf{F}(t)$ can be decomposed in the bases
 \begin{equation}\label{Ls}
 L_{1}=\left(
             \begin{array}{ccc}
               0 & 0 & 0 \\
               0 & 0 & -1 \\
               0 & 1 & 0 \\
             \end{array}
           \right),
 \quad
 L_{2}=\left(
             \begin{array}{ccc}
               0 & 0 & 1 \\
               0 & 0 & 0 \\
               -1 & 0 & 0 \\
             \end{array}
           \right),
\quad
L_{3}=\left(
             \begin{array}{ccc}
               0 & -1 & 0 \\
               1 & 0 & 0 \\
               0 & 0 & 0 \\
             \end{array}
           \right).
\end{equation}
These matrices are linearly independent and they satisfy the commutation relation $ [\mathbf{L}_{i},\mathbf{L}_{j}]=\varepsilon_{ijk}\mathbf{L}_{k}$
in which $\varepsilon_{ijk} $ is the Levi-Civita symbol. From iteration method, one finds the solution to (\ref{udotf}) as
\begin{eqnarray}\label{Q3}
\mathbf{\hat{u}}(t) &=& \mathbf{\hat{u}}(0)+\sum_{n=1}^{\infty}\int_{0}^{t_{1}}dt_{2}\int_{0}^{t_{2}}dt_{3}\cdots\int_{0}^{t_{n}}dt_{n-1}dt_{n}\nonumber\\
&\times& \mathbf{F}(t_{1})\cdot\mathbf{F}(t_{2})\cdots\mathbf{F}(t_{n})\cdot\mathbf{\hat{u}}(0).
\end{eqnarray}
In order to find the mean square rotational displacement one has to calculate
\begin{equation}\label{msr}
  \langle\Delta\mathbf{\hat{u}}^{2}(t)\rangle=\langle |\mathbf{\hat{u}}(t)-\mathbf{\hat{u}}(0)|^{2} \rangle=2[1-\langle \mathbf{\hat{u}}(t)\rangle\cdot\mathbf{\hat{u}}(0)],
\end{equation}
where the mean velocity is
\begin{eqnarray}\label{umean}
&& \langle\mathbf{\hat{u}}(t)\rangle=\sum_{n=1}^{\infty}
\int_{0}^{t_{1}}dt_{2}\int_{0}^{t_{2}}dt_{3}\cdots\int_{0}^{t_{n}}dt_{n-1}dt_{n}\nonumber\\
&\times& \langle\mathbf{F}(t_{1})\cdot\mathbf{F}(t_{2})\cdots\mathbf{F}(t_{n})\rangle\cdot\mathbf{\hat{u}}(0).
\end{eqnarray}
In $L_{i}$ bases the matrix $\mathbf{F}(t)$ can be written as
\begin{equation}\label{FL}
  \mathbf{F}(t)=\mathbf{T}_{1}(t)\mathbf{L}_{1}+\mathbf{T}_{2}(t)\mathbf{L}_{2}+\mathbf{T}_{3}(t)\mathbf{L}_{3}.
\end{equation}
Using the Wick's theorem one can find the $n$-pint correlation functions easily. In high temperature regime (classical limit) the $\mathbf{T}(t)$ functions are delta-correlated at different times. Applying this property and also the identity $ \mathbf{L}_{i}^{2}=1 $ one finds
\begin{equation}\label{FF}
\langle\mathbf{F}(t_{1})\cdot\mathbf{F}(t_{2})\rangle=\sum_{k,j}\delta_{kj}\mathbf{L}_{k}\mathbf{L}_{j}\delta(t_{1}-t_{2})=-4\gamma_{r}K_{B}T\delta(t_{1}-t_{2}),
\end{equation}
now from Wick's theorem and (\ref{FF},\ref{umean}) we find
\begin{equation}\label{Q4}
\mathbf{\hat{u}}(t)=\sum_{n=1}^{\infty}(\frac{1}{2})^{n}(-4\gamma_{r}K_{B}T)^{n/2}\frac{t^{n/2}}{(n/2)!}=e^{-2D_{r}t}\mathbf{\hat{u}}(0),
\end{equation}
where $ D_{r}=k_{B}T/\gamma_{r}.$ Therefore,
\begin{equation}\label{Du2}
\langle \Delta\mathbf{u}(t)^{2} \rangle=2[1-exp(-2D_{r}t)].
\end{equation}
\section{conclusion}
By modeling the environment by two independent vector fields the quantum Brownian motion of a rod-like particle was investigated in the frame work of canonical quantization. The quantum mechanical and classical limits for both Translational and rotational motions were obtained. Explicit relations for correlation functions, fluctuation-dissipation relations and mean squared values for both translational and rotational motions were derived.

\end{document}